\documentclass[a4paper, 10pt, twocolumn]{article} 
\usepackage[british]{babel}
\usepackage{microtype}
\usepackage{amsmath,amsfonts,amsthm}
\usepackage[svgnames,table]{xcolor}
\definecolor{GammaDarkGreen}{rgb}{0.113,0.278,0.2745}
\definecolor{GammaCharcoal}{rgb}{0.0706, 0.0706, 0.0706}
\definecolor{GammaRed}{HTML}{B71325}
\definecolor{GammaGreen}{HTML}{004846}
\usepackage{booktabs}
\usepackage{lastpage}
\usepackage{graphicx}
\usepackage{enumitem}
\setlist{leftmargin=*,itemsep=0.4ex,parsep=0ex,topsep=0.8ex,partopsep=0ex}
\setlist[itemize]{label=\color{GammaRed}$\triangleright$}
\usepackage{sectsty}
\chapterfont{\normalfont\Large\color{GammaRed}}
\sectionfont{\normalfont\Large\color{GammaRed}}
\subsectionfont{\normalfont\large\color{GammaCharcoal}}
\subsubsectionfont{\normalfont\normalsize\color{GammaGreen}}

\usepackage{subfig}

\usepackage{caption}
\DeclareCaptionLabelFormat{abbrevfig}{\color{GammaRed}Fig.\ #2}
\DeclareCaptionLabelFormat{abbrevtab}{\color{GammaRed}Tab.\ #2}
\usepackage[pdftex]{hyperref}
\pdftrailerid{}
\hypersetup{colorlinks=true,linkcolor=GammaDarkGreen,citecolor=GammaDarkGreen,filecolor=GammaDarkGreen,urlcolor=GammaDarkGreen}

\addto\extrasenglish{%
}

\usepackage{geometry}
\usepackage[utf8]{inputenc}
\usepackage[T1]{fontenc}
\IfFileExists{nimbusmononarrow.sty}{\usepackage{nimbusmononarrow}}{}
\usepackage{tgheros}
\usepackage{sansmath}
\usepackage{fancyhdr}
\usepackage{titling}

\newcommand{\authorstyle}[1]{{\large\color{GammaCharcoal}#1}}
\newcommand{\institution}[1]{{\footnotesize\itshape\color{GammaDarkGreen}#1\/}}

\pretitle{%
    \noindent
    \begin{minipage}[t]{0.63\textwidth}\vspace{20pt}
        \LARGE\color{GammaRed}
}

\posttitle{%
    \end{minipage}%
    \hfill
    \begin{minipage}[t]{0.33\textwidth}\vspace{-50pt}
        \raggedleft
        \includegraphics[width=\linewidth]{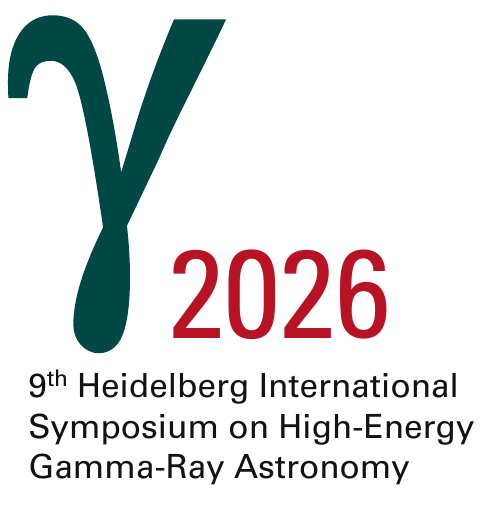}
    \end{minipage}%
}

\preauthor{\color{GammaDarkGreen}}
\postauthor{}
\predate{\begin{flushleft}\large\color{GammaRed}}
\postdate{\par\end{flushleft}}

\usepackage{lettrine}
\usepackage{xstring}
\newcommand{\initial}[1]{\lettrine[lines=2,findent=4pt,nindent=0pt]{\color{GammaRed}\emph{#1}}{}}
\newcommand{\lettrineabstract}[1]{\StrLeft{#1}{1}[\firstletter]\initial{\firstletter}\emph{{\StrGobbleLeft{#1}{1}}}}

\usepackage[backend=bibtex,style=numeric-comp,sorting=none,giveninits=true,citestyle=numeric]{biblatex}
\usepackage[autostyle=true]{csquotes}

\usepackage{aas_macros}

\DeclareUnicodeCharacter{2024}{\textperiodcentered}

\newcommand{\exhale}{\textit{ExHaLe-jet}}
\newcommand{\g}{\ensuremath{\gamma}}
\newcommand{\std}{\textbf{Std}}
\newcommand{\lowedd}{\textbf{LowEdd}}
\newcommand{\distacc}{\textbf{DistAcc}}

\title{
    \textit{ExHaLe-jet}: Modeling blazar jets with an extended hadro-leptonic radiation code 
}

\author{
	\authorstyle{M.~Zacharias\textcolor{GammaDarkGreen}{\textsuperscript{1}}, A. Zech\textcolor{GammaDarkGreen}{\textsuperscript{2}}, A. Reimer\textcolor{GammaDarkGreen}{\textsuperscript{3}}, C. Boisson\textcolor{GammaDarkGreen}{\textsuperscript{2}}}
	\newline\newline
    {\footnotesize
        \textsuperscript{\color{GammaDarkGreen}1}\institution{Landessternwarte, Universit\"at Heidelberg, K\"onigstuhl 12, D-69117 Heidelberg, Germany}\\
        \textsuperscript{\color{GammaDarkGreen}2}\institution{LUX, Observatoire de Paris - PSL, CNRS, SU, Paris, France}\\
        \textsuperscript{\color{GammaDarkGreen}3}\institution{ Universit\"at Innsbruck, Institut f\"ur Astro- und Teilchenphysik, Technikerstr. 25, A-6020 Innsbruck, Austria}
    }\\
}

\date{\today}
\newcommand{\website}{\href{https://plan.events.mpg.de/event/543/}{Proceedings of the 9th Heidelberg Gamma-Ray Symposium}} 

\begin{document}
\sansmath\maketitle
\thispagestyle{fancy}

\lettrineabstract{Blazars – active galaxies with the jet pointing at Earth – emit across all electromagnetic wavelengths. As the one-zone model becomes increasingly inadequate to describe blazar emission, more elaborate models are needed. In order to self-consistently describe the entire electromagnetic spectrum, extended jet models are necessary. Notably, kinetic descriptions of extended jets can provide the temporal and spatial evolution of the particle species and the full electromagnetic output. {ExHaLe-jet} is a lepto-hadronic, extended-jet code allowing for in-depth studies of the particle and radiation evolution on all scales.
Here, we present our ongoing study regarding acceleration scenarios of particles within the large-scale jet structure. A single acceleration of particles at the jet base (or elsewhere) is insufficient to produce sufficient photon flux. Thus, re-acceleration through, e.g., consecutive shocks along the jet is needed. We explore two possible scenarios for this.
}

\section{Introduction}

\exhale\ is the first kinetic radiation code that describes the radiation output of relativistic jets from their base up to their end using both leptonic and hadronic processes \cite{zacharias+22}. Such a code becomes increasingly important due to the obvious limitations of the one-zone model.

Relativistic jets -- with blazars as a prominent example -- are among the most energetically extreme environments in the Universe. They are relativistically moving entities containing particles, which are themselves accelerated to relativistic energies in the comoving frame of the jet. In turn, jets can be extremely luminous with rapid and powerful flares -- with the most extreme ones belonging to \g-ray bursts. 

Jets emit radiation across the entire electromagnetic spectrum. In most cases, the spectral energy distribution (SED) shows two broad components, where the low-energy one is attributed to synchrotron emission of electrons (incl. positrons). The high-energy component is generally assumed to be inverse-Compton emission of electrons scattering ambient photon fields such as synchrotron (the so-called synchrotron-self-Compton process, SSC) from within the jet, or thermal fields from outside the jet (the so-called external-Compton process) including the accretion disk, the broad-line region, the dusty torus or the cosmic microwave background \cite{cerruti20}. However, if the association of neutrinos with jets is confirmed \cite{icecube+18}, the presence of relativistic protons is highly probable, enabling additional processes to produce \g\ rays, such as proton synchrotron or secondary cascades \cite{boettcher+13}.

The rapid, and in many cases correlated flares have given rise to the one-zone model, where practically the entire electromagnetic output is attributed to a small region somewhere in the larger jet structure. While this can be used to explain (some) flares \cite{zacharias+17,hess19,hess26_0903}, it already faces problems with others \cite{hess+21}, and is most certainly inadequate to explain the long-term evolution of jets \cite{Taylor+26,WZ26,ZW26}. Hence, multi-zone and extended-jet models, such as \exhale, are required to properly describe the emission of relativistic jets.

In the following, we first describe the basic functionalities of the code with all details in \cite{zacharias+22}, followed by the description of the setup used for our study. Section~4 describes the results, after which we present our conclusions.

\section{Summary of the code}

The jet is modelled as an initially parabolic region, where the bulk flow accelerates. Once the jet reaches its maximum bulk Lorentz factor at the end of the bulk-acceleration zone, it becomes conical with constant speed. This geometry is inspired by several observations \cite{boccardi+21,park+21}.

Along its axis, the jet is cut into numerous slices with equal logarithmic spacing resulting in the same resolution per distance decade. In each slice, a time-dependent Fokker-Planck equation is solved including injection and acceleration terms, cooling, and particle decay. We consider protons, pions, muons, and electrons. We assume that particles only move downstream in the jet, but cannot leave through the side. Simultaneously with the particle equation, we also solve the radiation transport equation to obtain the radiative output of the given slice, which also feeds back to the Fokker-Planck equation for, e.g., absorption processes. In each slice, the neutrino yield is also derived.

An initial injection of protons and electrons takes place in the first slice. Once the solution to the Fokker-Planck equation is obtained, the resulting particle distribution is used as the injection function in the next slice; modulo a change in density due to the expansion and possible change in bulk speed. Secondaries also move along. Hence, if no additional particle acceleration takes place, particles will mostly cool as they flow along the jet. Re-acceleration in specific slices can be modelled by using only the updated density from the previous slice but otherwise set a new injection form for the primary particles (protons and electrons). This allows us to test various acceleration scenarios along the jet flow; one of which is shown here.

Next to the intrinsically produced photon fields, the code also considers external photon fields, namely the accretion disk, the broad-line region, the dusty torus, and the cosmic microwave background. Except for the latter, these fields are distance-dependent. The luminosity and size of the broad-line region and the dusty torus are scaled by the accretion disk luminosity, while their temperatures are free parameters. Hence, a lower accretion disk luminosity implies both lower luminosities and smaller radii for the broad-line region and the dusty torus.

\section{Simulation setup}

We try to recreate a flow, where the particles are re-accelerated at several shocks along the jet flow. In the standard setup (hereafter \std), we reshape the particle distribution into an acceleration spectrum at the base and then at 9 shocks placed at equal logarithmic distance. In each shock, we set the same particle acceleration spectrum for the protons and electrons irrespective of position in the jet and local magnetic field. The only parameter that changes in the acceleration/injection spectrum are the local particle densities. This arbitrary setup allows us to study the impact of the various shocks on the total emitted photon flux, as well as the relative importance of their location. The free parameters used for the \std\ simulation are given in Tab.~\ref{tab:freepara}.

\begin{table}
    \centering
    \begin{tabular}{|p{4.7cm}|p{2.5cm}|}
        \hline
        \rowcolor{lightgray}
        \large{\textbf{Parameter}} & \large{\textbf{Value}}  \\
        \hline \hline
        Jet length & $100\,$pc \\
        \hline
        Bulk flow acceleration length & {$1\,$pc} \\
        \hline
        Maximum bulk flow Lorentz factor & {$30$} \\
        \hline
        Magnetic field strength at jet base & {$200\,$G} \\
        \hline
        Injection particle power at jet base & {$2.2\times 10^{-4}\,L_{\rm edd}$} \\
        \hline
        Injection electron to proton ratio at jet base & {$15$} \\
        \hline
        Acceleration proton minimum/maximum Lorentz factors & {$2.0$ / $2.0\times 10^{8}$} \\
        \hline
        Acceleration proton spectral index & {$2.5$} \\
        \hline
        Acceleration electron minimum/maximum Lorentz factors & {$100$ / $1.0\times 10^{5}$} \\
        \hline
        Acceleration electron spectral index & {$2.5$} \\
        \hline
        Accretion disk luminosity & $0.1\,L_{\rm edd}$ \\
        \hline
        Broad-line region temperature &  {$10^4\,$K} \\
        \hline
        Dusty Torus temperature &  {$500\,$K} \\
        \hline
    \end{tabular}
    \caption{Parameter setting for the \std\ simulation, where $L_{\rm edd}$ is the Eddington luminosity of the black hole with mass $3\times 10^8\,M_{\odot}$.}
    \label{tab:freepara}
\end{table}

We also perform two additional simulations. In the first one, we only change the accretion disk luminosity to $10^{-3}\,L_{\rm edd}$. This simulation is referred to as \lowedd. The other simulation has the same free parameters as the \std\ simulation, but next to particle acceleration at the base, we only place further shocks where the magnetization of the jet is less than unity -- which is the case in the conical section of the shock. This reduces the number of shocks from 10 to 4. This simulation is referred to as \distacc.

\section{Results}

\begin{figure*}
    \begin{minipage}[h]{0.65\textwidth}
        \includegraphics[width=0.98\textwidth]{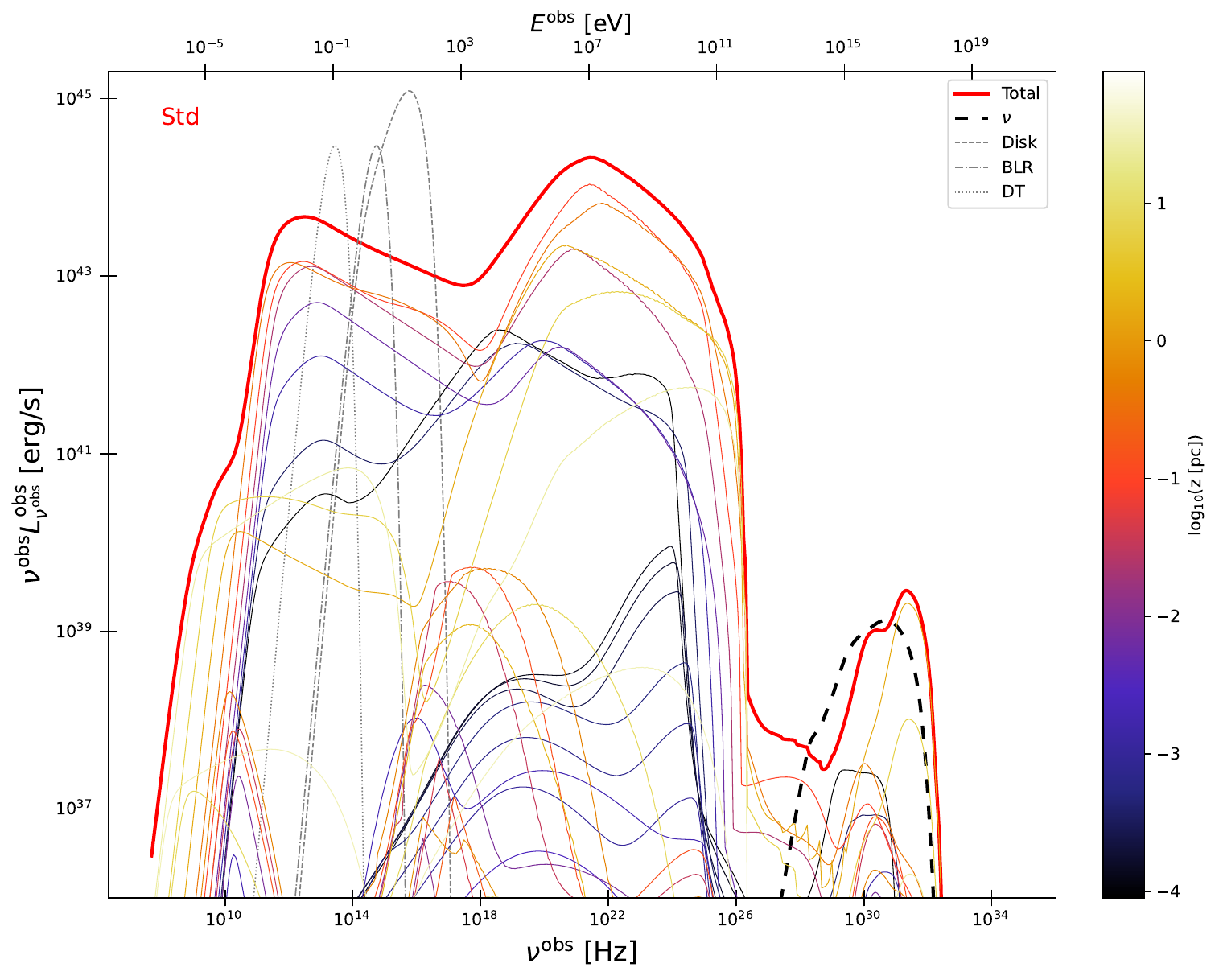}
    \end{minipage}~
    \begin{minipage}[h]{0.33\textwidth}
        \includegraphics[width=0.97\textwidth]{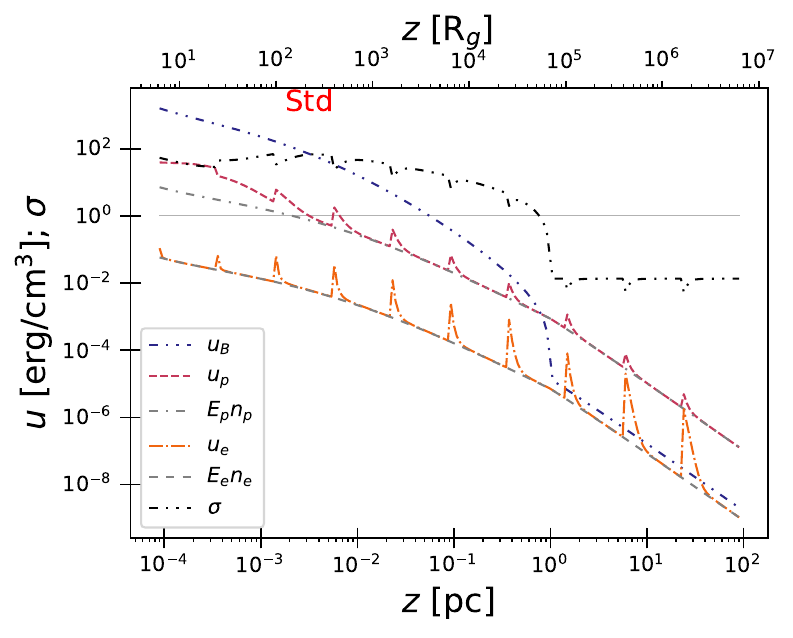}\\
        \includegraphics[width=0.97\textwidth]{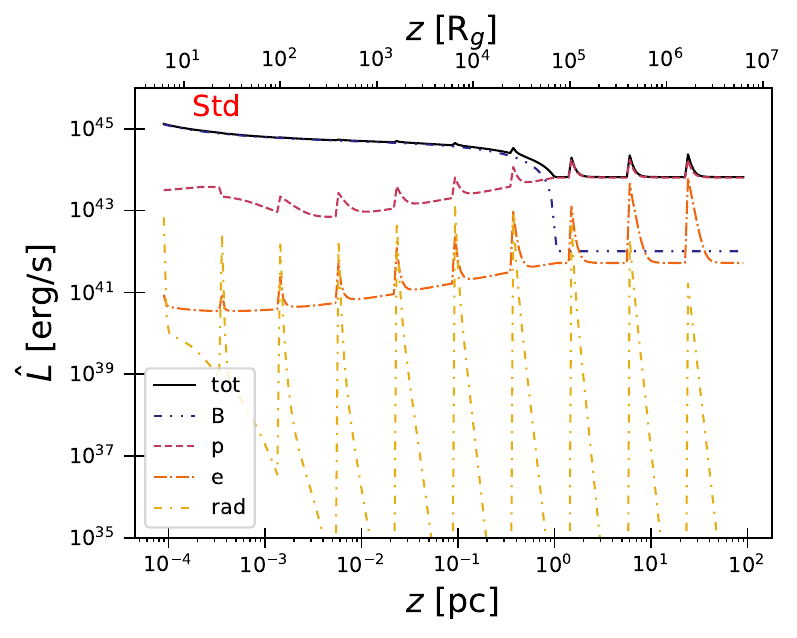}
    \end{minipage}
    \caption{Results of the \textbf{Std} simulation. \textit{Left:} Distance evolution (color code) of the intrinsic SED. The total is given by the thick red line. The black dashed line marks the total neutrino spectrum, while the gray lines mark the external fields as indicated. \textit{Right:} Evolution of the energy densities and the magnetization in the comoving frame (top) and of the luminosities in the observer's frame (bottom) as a function of jet length for various constituents as indicated. All entities are derived at the end of the calculation of a given slice (aka, they include cooling effects, etc.).}
    \label{fig:std}
\end{figure*}

\begin{figure*}
    \begin{minipage}[h]{0.65\textwidth}
        \includegraphics[width=0.98\textwidth]{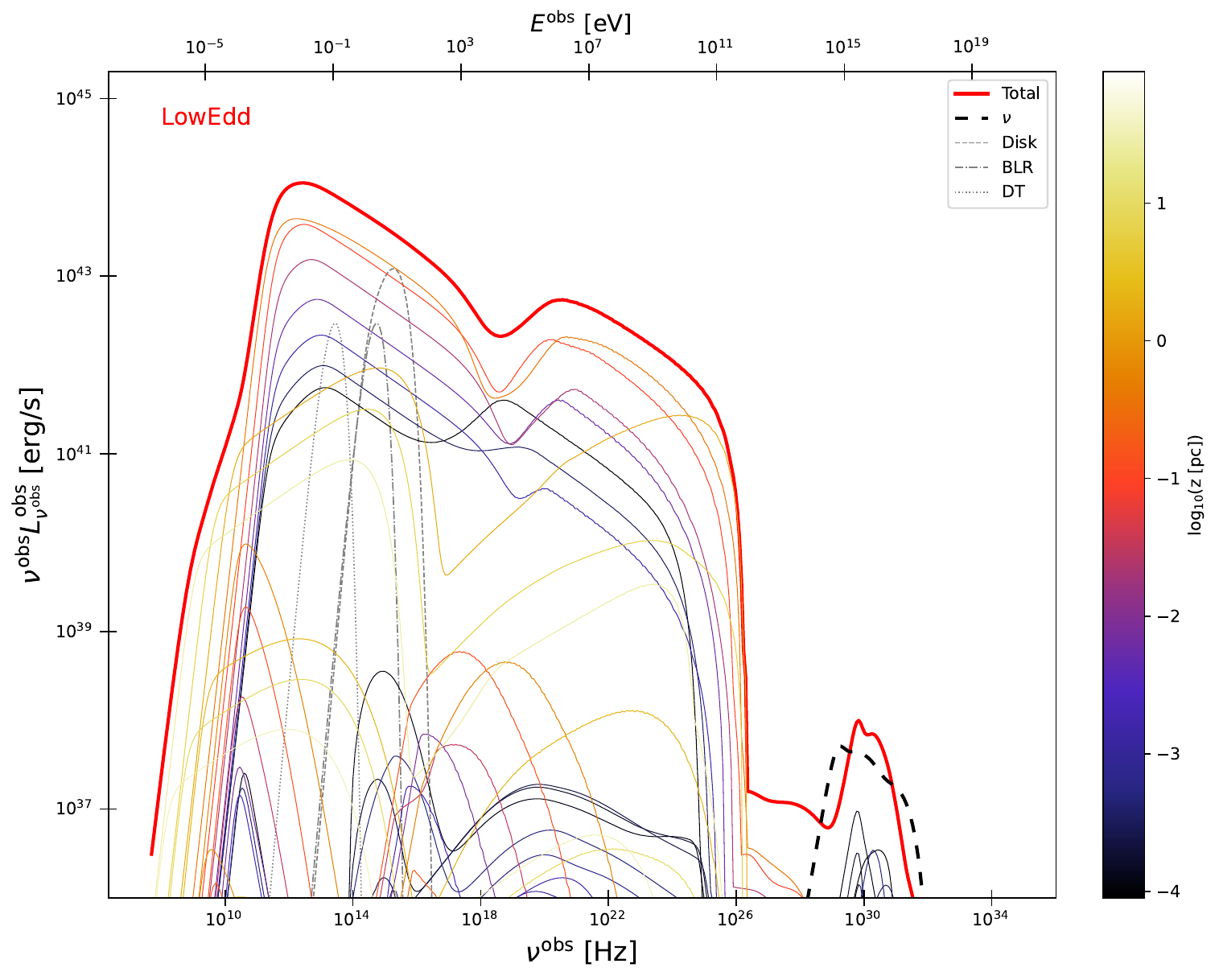}
    \end{minipage}~
    \begin{minipage}[h]{0.33\textwidth}
        \includegraphics[width=0.97\textwidth]{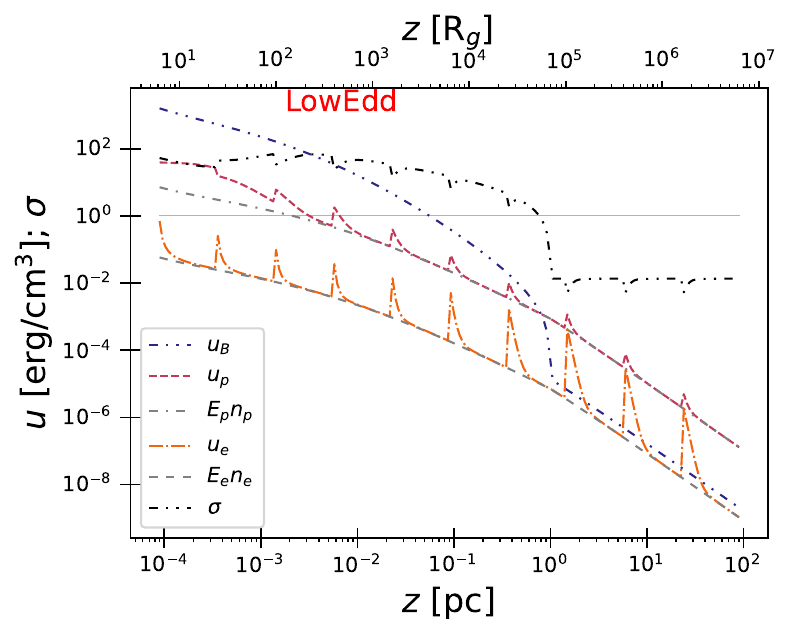}\\
        \includegraphics[width=0.97\textwidth]{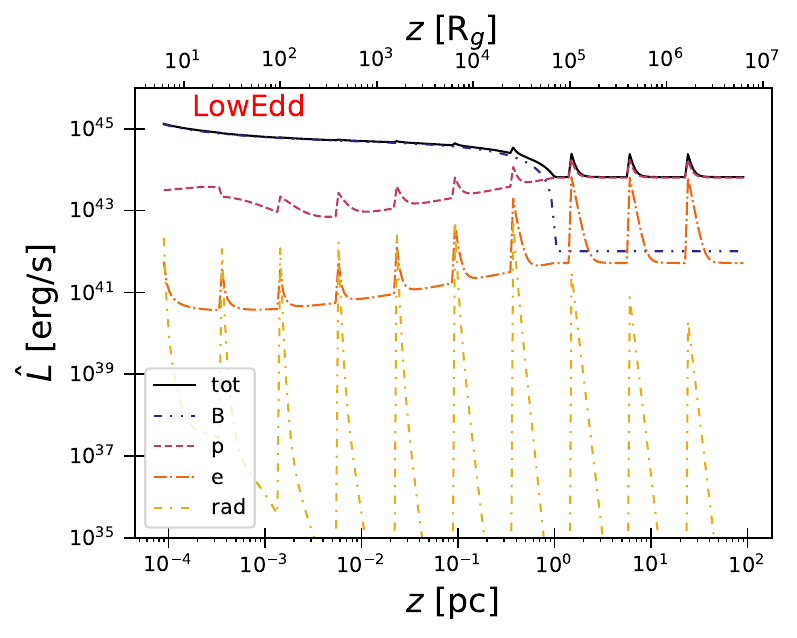}
    \end{minipage}
    \caption{Same as Fig.~\ref{fig:std}, but for the \textbf{LowEdd} simulation.}
    \label{fig:lowedd}
\end{figure*}

\begin{figure*}
    \begin{minipage}[h]{0.65\textwidth}
        \includegraphics[width=0.98\textwidth]{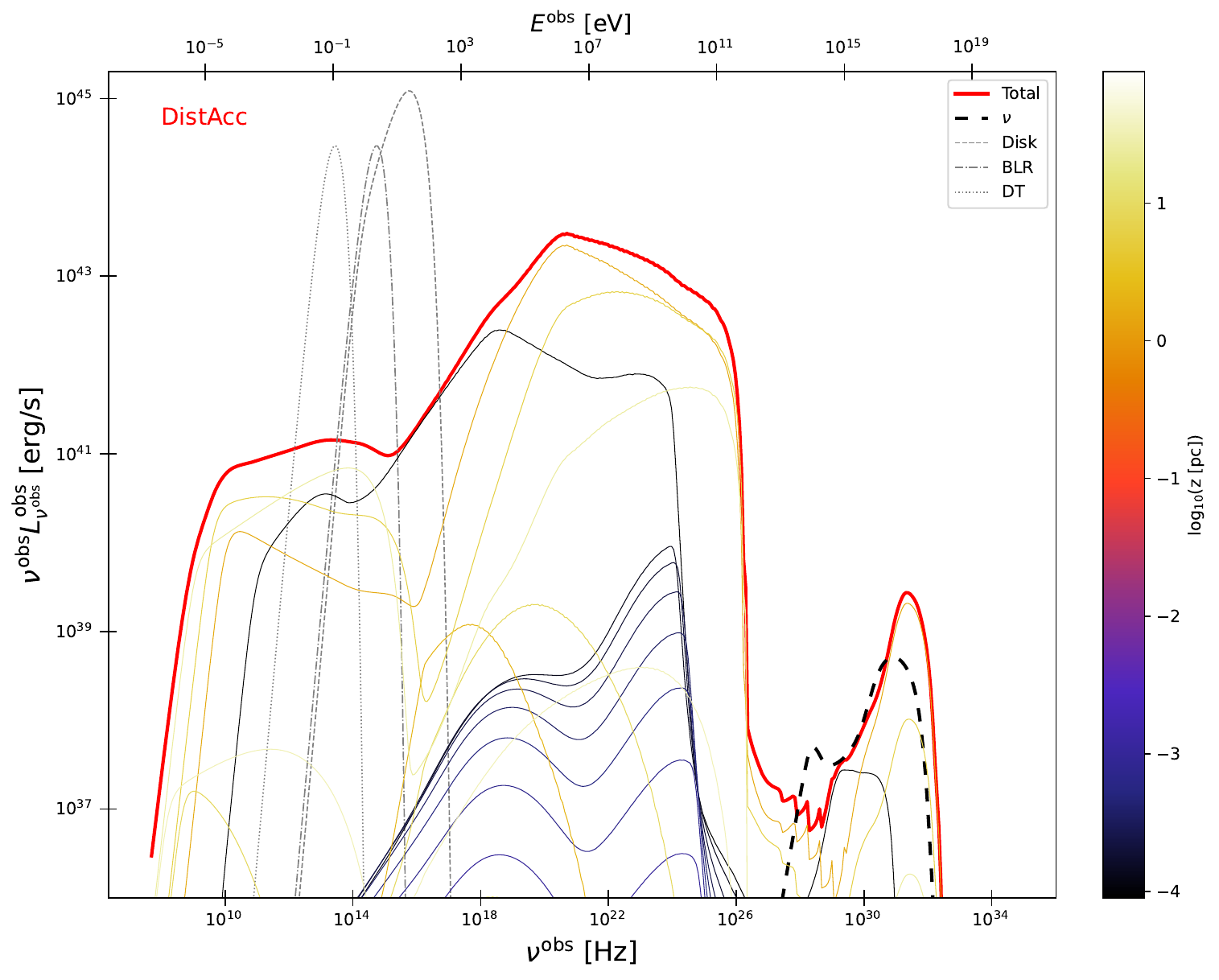}
    \end{minipage}~
    \begin{minipage}[h]{0.33\textwidth}
        \includegraphics[width=0.97\textwidth]{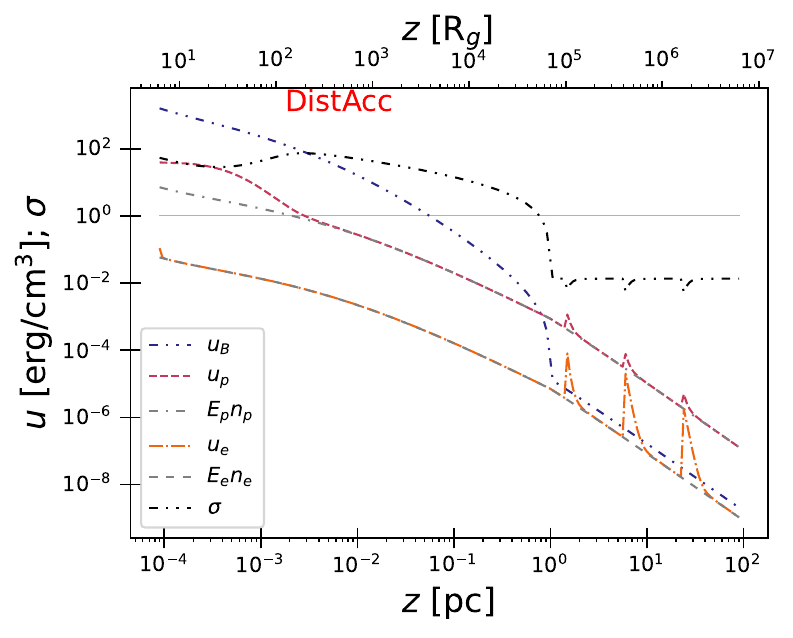}\\
        \includegraphics[width=0.97\textwidth]{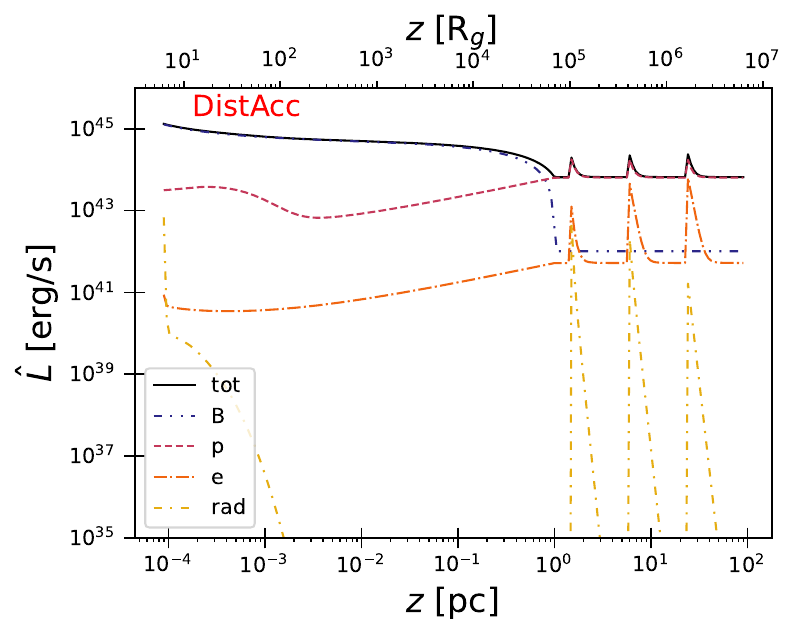}
    \end{minipage}
    \caption{Same as Fig.~\ref{fig:std}, but for the \textbf{DistAcc} simulation.}
    \label{fig:distacc}
\end{figure*}

The results for the 3 simulations are shown in Figs.~\ref{fig:std}-\ref{fig:distacc}. Each figure shows the SED and its evolution as a function of distance along the jet, as well as the evolution of the energy densities in the comoving frame and the evolution of the luminosities in the observer's frame -- both also as a function of distance along the jet. The energy densities and luminosities are given for different jet constituents, namely the magnetic field, protons, electrons, and radiation. The energy-density plot also contains the magnetization, $\sigma$, of the jet.

The SED in the \std\ simulation is dominated by the inverse-Compton component, which is mainly due to scattering of broad-line-region and dusty-torus photons. It is evident that only the small number of slices at and shortly after each shock actually contribute meaningfully to the total flux. The strongest contributions come from slices around the end of the bulk acceleration, where beaming is close to or even at maximum, while the energy densities have not dropped too much yet. After each shock, the particles cool fairly quickly, as can be seen by the energy density $u$ of the particles dropping down to the $E\,n$ lines, which are the particles rest mass energy, $E$, multiplied with the particle number density, $n$. It appears that, especially for electrons, the cooling distance increases considerable with larger distance from the black hole. This clearly demonstrates the reduced cooling rate due to the decrease in the magnetic field and the external photon fields. For protons, there is not such an obvious difference in the cooling length with increasing distance, which may be a sign that the proton cooling is not dominated by interactions with these fields, but rather by the adiabatic expansion.

The effect of reducing the accretion disk luminosity, which implies reduced luminosity and size of the broad-line region and the dusty torus, is clearly visible in Fig.~\ref{fig:lowedd}, which shows the results of the \lowedd\ simulation. In this case, the synchrotron component dominates the SED, while the external-Compton component is much reduced. The reduction in the $\pi^0$ and neutrino bumps ($\nu^{\rm obs}\sim10^{30}\,$Hz) compared to the \std\ simulation also indicate reductions in p-\g\ interactions.

Letting instead particles only reaccelerate once $\sigma<1$ (\distacc\ simulation, Fig.~\ref{fig:distacc}), the reduced magnetic field precludes any substantial synchrotron production at these distances. As the external photon fields are still strong, the \g-ray production is only mildly reduced compared to the \std\ simulation due to the missing contribution towards the end of the bulk-acceleration zone.

\section{Conclusion}

We have studied the effect of several shocks located at various positions within a blazar jet using the extended, hadro-leptonic radiation code \exhale\ \cite{zacharias+22}. Our results can be summarized as follows. 

It is obvious that particles must be reaccelerated several times within a jet, as a single reacceleration is not sufficient to provide sufficient photon flux. While Doppler beaming is helpful to boost the received flux, a sufficiently high magnetic field and external photon fields are also important. The latter suggests that re-acceleration points should be located at best around the end of the bulk-acceleration zone -- however, on both sides of it in order to benefit from both high magnetic fields and high Doppler factors, respectively. Despite the fact of relativistic protons being present, no observable neutrino flux is produced in either of our simulations.

Here, we have only covered a single scenario for reacceleration of particles. However, potentially more continuous processes, such as magnetic reconnection \cite{zhang+21}, turbulence \cite{zechlemoine21,tavecchio+22} or shear layers \cite{rieger19} could produce extended acceleration zones with corresponding continuously high photon emission. This will be discussed in our forthcoming publication.

\printbibliography[title={References}]

@ARTICLE{zacharias+22,
       author = {{Zacharias}, M. and {Reimer}, A. and {Boisson}, C. and {Zech}, A.},
        title = "{EXHALE-JET: an extended hadro-leptonic jet model for blazars - I. Code description and initial results}",
      journal = {\mnras},
         year = 2022,
        month = may,
       volume = {512},
       number = {3},
        pages = {3948-3971},
          doi = {10.1093/mnras/stac754},
archivePrefix = {arXiv},
       eprint = {2203.07956},
 primaryClass = {astro-ph.HE},
       adsurl = {https://ui.adsabs.harvard.edu/abs/2022MNRAS.512.3948Z}
}

@ARTICLE{boettcher+13,
       author = {{B{\"o}ttcher}, M. and {Reimer}, A. and {Sweeney}, K. and {Prakash}, A.},
        title = "{Leptonic and Hadronic Modeling of Fermi-detected Blazars}",
      journal = {\apj},
         year = 2013,
        month = may,
       volume = {768},
       number = {1},
          eid = {54},
        pages = {54},
          doi = {10.1088/0004-637X/768/1/54},
archivePrefix = {arXiv},
       eprint = {1304.0605},
 primaryClass = {astro-ph.HE},
       adsurl = {https://ui.adsabs.harvard.edu/abs/2013ApJ...768...54B}
}

@ARTICLE{zacharias+17,
       author = {{Zacharias}, M. and {B{\"o}ttcher}, M. and {Jankowsky}, F. and {Lenain}, J. -P. and {Wagner}, S.~J. and {Wierzcholska}, A.},
        title = "{Cloud Ablation by a Relativistic Jet and the Extended Flare in CTA 102 in 2016 and 2017}",
      journal = {\apj},
         year = 2017,
        month = dec,
       volume = {851},
       number = {2},
          eid = {72},
        pages = {72},
          doi = {10.3847/1538-4357/aa9bee},
archivePrefix = {arXiv},
       eprint = {1711.06117},
 primaryClass = {astro-ph.HE},
       adsurl = {https://ui.adsabs.harvard.edu/abs/2017ApJ...851...72Z}
}

@ARTICLE{hess19,
       author = {{H.E.S.S. Collaboration} and {Abdalla}, H. and {Adam}, R. and {Aharonian}, F. and {Ait Benkhali}, F. and {Ang{\"u}ner}, E.~O. and {Arakawa}, M. and {Arcaro}, C. and {Armand}, C. and {Ashkar}, H. and {Backes}, M. and {Barbosa Martins}, V. and {Barnard}, M. and {Becherini}, Y. and {Berge}, D. and {Bernl{\"o}hr}, K. and {Blackwell}, R. and {B{\"o}ttcher}, M. and {Boisson}, C. and {Bolmont}, J. and {Bonnefoy}, S. and {Bregeon}, J. and {Breuhaus}, M. and {Brun}, F. and {Brun}, P. and {Bryan}, M. and {B{\"u}chele}, M. and {Bulik}, T. and {Bylund}, T. and {Capasso}, M. and {Caroff}, S. and {Carosi}, A. and {Casanova}, S. and {Cerruti}, M. and {Chand}, T. and {Chandra}, S. and {Chen}, A. and {Colafrancesco}, S. and {Cury{\l}o}, M. and {Davids}, I.~D. and {Deil}, C. and {Devin}, J. and {deWilt}, P. and {Dirson}, L. and {Djannati-Ata{\"\i}}, A. and {Dmytriiev}, A. and {Donath}, A. and {Doroshenko}, V. and {Drury}, L.~O. 'C. and {Dyks}, J. and {Egberts}, K. and {Emery}, G. and {Ernenwein}, J. -P. and {Eschbach}, S. and {Feijen}, K. and {Fegan}, S. and {Fiasson}, A. and {Fontaine}, G. and {Funk}, S. and {F{\"u}{\ss}ling}, M. and {Gabici}, S. and {Gallant}, Y.~A. and {Gat{\'e}}, F. and {Giavitto}, G. and {Glawion}, D. and {Glicenstein}, J.~F. and {Gottschall}, D. and {Grondin}, M. -H. and {Hahn}, J. and {Haupt}, M. and {Heinzelmann}, G. and {Henri}, G. and {Hermann}, G. and {Hinton}, J.~A. and {Hofmann}, W. and {Hoischen}, C. and {Holch}, T.~L. and {Holler}, M. and {Horns}, D. and {Huber}, D. and {Iwasaki}, H. and {Jamrozy}, M. and {Jankowsky}, D. and {Jankowsky}, F. and {Jardin-Blicq}, A. and {Jung-Richardt}, I. and {Kastendieck}, M.~A. and {Katarzy{\'n}ski}, K. and {Katsuragawa}, M. and {Katz}, U. and {Khangulyan}, D. and {Kh{\'e}lifi}, B. and {King}, J. and {Klepser}, S. and {Klu{\'z}niak}, W. and {Komin}, Nu. and {Kosack}, K. and {Kostunin}, D. and {Kraus}, M. and {Lamanna}, G. and {Lau}, J. and {Lemi{\`e}re}, A. and {Lemoine-Goumard}, M. and {Lenain}, J. -P. and {Leser}, E. and {Levy}, C. and {Lohse}, T. and {Lypova}, I. and {Mackey}, J. and {Majumdar}, J. and {Malyshev}, D. and {Marandon}, V. and {Marcowith}, A. and {Mares}, A. and {Mariaud}, C. and {Mart{\'\i}-Devesa}, G. and {Marx}, R. and {Maurin}, G. and {Meintjes}, P.~J. and {Mitchell}, A.~M.~W. and {Moderski}, R. and {Mohamed}, M. and {Mohrmann}, L. and {Moore}, C. and {Moulin}, E. and {Muller}, J. and {Murach}, T. and {Nakashima}, S. and {de Naurois}, M. and {Ndiyavala}, H. and {Niederwanger}, F. and {Niemiec}, J. and {Oakes}, L. and {O'Brien}, P. and {Odaka}, H. and {Ohm}, S. and {de Ona Wilhelmi}, E. and {Ostrowski}, M. and {Oya}, I. and {Panter}, M. and {Parsons}, R.~D. and {Perennes}, C. and {Petrucci}, P. -O. and {Peyaud}, B. and {Piel}, Q. and {Pita}, S. and {Poireau}, V. and {Priyana Noel}, A. and {Prokhorov}, D.~A. and {Prokoph}, H. and {P{\"u}hlhofer}, G. and {Punch}, M. and {Quirrenbach}, A. and {Raab}, S. and {Rauth}, R. and {Reimer}, A. and {Reimer}, O. and {Remy}, Q. and {Renaud}, M. and {Rieger}, F. and {Rinchiuso}, L. and {Romoli}, C. and {Rowell}, G. and {Rudak}, B. and {Ruiz-Velasco}, E. and {Sahakian}, V. and {Saito}, S. and {Sanchez}, D.~A. and {Santangelo}, A. and {Sasaki}, M. and {Schlickeiser}, R. and {Sch{\"u}ssler}, F. and {Schulz}, A. and {Schutte}, H. and {Schwanke}, U. and {Schwemmer}, S. and {Seglar-Arroyo}, M. and {Senniappan}, M. and {Seyffert}, A.~S. and {Shafi}, N. and {Shiningayamwe}, K. and {Simoni}, R. and {Sinha}, A. and {Sol}, H. and {Specovius}, A. and {Spir-Jacob}, M. and {Stawarz}, L. and {Steenkamp}, R. and {Stegmann}, C. and {Steppa}, C. and {Takahashi}, T. and {Tavernier}, T. and {Taylor}, A.~M. and {Terrier}, R. and {Tiziani}, D. and {Tluczykont}, M. and {Trichard}, C. and {Tsirou}, M. and {Tsuji}, N. and {Tuffs}, R. and {Uchiyama}, Y. and {van der Walt}, D.~J. and {van Eldik}, C. and {van Rensburg}, C. and {van Soelen}, B. and {Vasileiadis}, G. and {Veh}, J. and {Venter}, C. and {Vincent}, P. and {Vink}, J. and {Voisin}, F. and {V{\"o}lk}, H.~J. and {Vuillaume}, T. and {Wadiasingh}, Z. and {Wagner}, S.~J. and {White}, R. and {Wierzcholska}, A. and {Yang}, R. and {Yoneda}, H. and {Zacharias}, M. and {Zanin}, R. and {Zdziarski}, A.~A. and {Zech}, A. and {Ziegler}, A. and {Zorn}, J. and {{\.Z}ywucka}, N. and {Meyer}, M.},
        title = "{Constraints on the emission region of 3C 279 during strong flares in 2014 and 2015 through VHE {\ensuremath{\gamma}}-ray observations with H.E.S.S.}",
      journal = {\aap},
         year = 2019,
        month = jul,
       volume = {627},
          eid = {A159},
        pages = {A159},
          doi = {10.1051/0004-6361/201935704},
archivePrefix = {arXiv},
       eprint = {1906.04996},
 primaryClass = {astro-ph.HE},
       adsurl = {https://ui.adsabs.harvard.edu/abs/2019A&A...627A.159H}
}

@ARTICLE{hess+21,
       author = {{H.E.S.S. Collaboration} and {Abdalla}, H. and {Adam}, R. and {Aharonian}, F. and {Ait Benkhali}, F. and {Ang{\"u}ner}, E.~O. and {Arcaro}, C. and {Armand}, C. and {Armstrong}, T. and {Ashkar}, H. and {Backes}, M. and {Baghmanyan}, V. and {Barbosa Martins}, V. and {Barnacka}, A. and {Barnard}, M. and {Becherini}, Y. and {Berge}, D. and {Bernl{\"o}hr}, K. and {Bi}, B. and {B{\"o}ttcher}, M. and {Boisson}, C. and {Bolmont}, J. and {Bonnefoy}, S. and {de Bony de Lavergne}, M. and {Bregeon}, J. and {Breuhaus}, M. and {Brun}, F. and {Brun}, P. and {Bryan}, M. and {B{\"u}chele}, M. and {Bulik}, T. and {Bylund}, T. and {Caroff}, S. and {Carosi}, A. and {Casanova}, S. and {Chand}, T. and {Chandra}, S. and {Chen}, A. and {Cotter}, G. and {Cury{\l}o}, M. and {Damascene Mbarubucyeye}, J. and {Davids}, I.~D. and {Davies}, J. and {Deil}, C. and {Devin}, J. and {Dewilt}, P. and {Dirson}, L. and {Djannati-Ata{\"\i}}, A. and {Dmytriiev}, A. and {Donath}, A. and {Doroshenko}, V. and {Dyks}, J. and {Egberts}, K. and {Eichhorn}, F. and {Einecke}, S. and {Emery}, G. and {Ernenwein}, J. -P. and {Feijen}, K. and {Fegan}, S. and {Fiasson}, A. and {Fichet de Clairfontaine}, G. and {Filipovic}, M. and {Fontaine}, G. and {Funk}, S. and {F{\"u}{\ss}ling}, M. and {Gabici}, S. and {Gallant}, Y.~A. and {Giavitto}, G. and {Giunti}, L. and {Glawion}, D. and {Glicenstein}, J.~F. and {Gottschall}, D. and {Grondin}, M. -H. and {Hahn}, J. and {Haupt}, M. and {Hermann}, G. and {Hinton}, J.~A. and {Hofmann}, W. and {Hoischen}, C. and {Holch}, T.~L. and {Holler}, M. and {H{\"o}rbe}, M. and {Horns}, D. and {Huber}, D. and {Jamrozy}, M. and {Jankowsky}, D. and {Jankowsky}, F. and {Jardin-Blicq}, A. and {Joshi}, V. and {Jung-Richardt}, I. and {Kastendieck}, M.~A. and {Katarzy{\'n}ski}, K. and {Katz}, U. and {Khangulyan}, D. and {Kh{\'e}lifi}, B. and {Klepser}, S. and {Klu{\'z}niak}, W. and {Komin}, Nu. and {Konno}, R. and {Kosack}, K. and {Kostunin}, D. and {Kreter}, M. and {Lamanna}, G. and {Lemi{\`e}re}, A. and {Lemoine-Goumard}, M. and {Lenain}, J. -P. and {Levy}, C. and {Lohse}, T. and {Lypova}, I. and {Mackey}, J. and {Majumdar}, J. and {Malyshev}, D. and {Malyshev}, D. and {Marandon}, V. and {Marchegiani}, P. and {Marcowith}, A. and {Mares}, A. and {Mart{\'\i}-Devesa}, G. and {Marx}, R. and {Maurin}, G. and {Meintjes}, P.~J. and {Meyer}, M. and {Mitchell}, A.~M.~W. and {Moderski}, R. and {Mohamed}, M. and {Mohrmann}, L. and {Montanari}, A. and {Moore}, C. and {Morris}, P. and {Moulin}, E. and {Muller}, J. and {Murach}, T. and {Nakashima}, K. and {Nayerhoda}, A. and {de Naurois}, M. and {Ndiyavala}, H. and {Niederwanger}, F. and {Niemiec}, J. and {Oakes}, L. and {O'Brien}, P. and {Odaka}, H. and {Ohm}, S. and {Olivera-Nieto}, L. and {de Ona Wilhelmi}, E. and {Ostrowski}, M. and {Panter}, M. and {Panny}, S. and {Parsons}, R.~D. and {Peron}, G. and {Peyaud}, B. and {Piel}, Q. and {Pita}, S. and {Poireau}, V. and {Priyana Noel}, A. and {Prokhorov}, D.~A. and {Prokoph}, H. and {P{\"u}hlhofer}, G. and {Punch}, M. and {Quirrenbach}, A. and {Raab}, S. and {Rauth}, R. and {Reichherzer}, P. and {Reimer}, A. and {Reimer}, O. and {Remy}, Q. and {Renaud}, M. and {Rieger}, F. and {Rinchiuso}, L. and {Romoli}, C. and {Rowell}, G. and {Rudak}, B. and {Ruiz-Velasco}, E. and {Sahakian}, V. and {Sailer}, S. and {Sanchez}, D.~A. and {Santangelo}, A. and {Sasaki}, M. and {Scalici}, M. and {Sch{\"u}ssler}, F. and {Schutte}, H.~M. and {Schwanke}, U. and {Schwemmer}, S. and {Seglar-Arroyo}, M. and {Senniappan}, M. and {Seyffert}, A.~S. and {Shafi}, N. and {Shiningayamwe}, K. and {Simoni}, R. and {Sinha}, A. and {Sol}, H. and {Specovius}, A. and {Spencer}, S. and {Spir-Jacob}, M. and {Stawarz}, {\L}. and {Sun}, L. and {Steenkamp}, R. and {Stegmann}, C. and {Steinmassl}, S. and {Steppa}, C. and {Takahashi}, T. and {Tavernier}, T. and {Taylor}, A.~M. and {Terrier}, R. and {Tiziani}, D. and {Tluczykont}, M. and {Tomankova}, L. and {Trichard}, C. and {Tsirou}, M. and {Tuffs}, R. and {Uchiyama}, Y. and {van der Walt}, D.~J. and {van Eldik}, C. and {van Rensburg}, C. and {van Soelen}, B. and {Vasileiadis}, G. and {Veh}, J. and {Venter}, C. and {Vincent}, P. and {Vink}, J. and {V{\"o}lk}, H.~J. and {Vuillaume}, T. and {Wadiasingh}, Z. and {Wagner}, S.~J. and {Watson}, J. and {Werner}, F. and {White}, R. and {Wierzcholska}, A. and {Wong}, Yu. W. and {Yusafzai}, A. and {Zacharias}, M. and {Zanin}, R. and {Zargaryan}, D. and {Zdziarski}, A.~A. and {Zech}, A. and {Zhu}, S.~J. and {Zorn}, J. and {Zouari}, S. and {{\.Z}ywucka}, N. and {MAGIC Collaboration} and {Acciari}, V.~A. and {Ansoldi}, S. and {Antonelli}, L.~A. and {Arbet Engels}, A. and {Asano}, K. and {Baack}, D. and {Babi{\'c}}, A. and {Baquero}, A. and {Barres de Almeida}, U. and {Barrio}, J.~A. and {Becerra Gonz{\'a}lez}, J. and {Bednarek}, W. and {Bellizzi}, L. and {Bernardini}, E. and {Berti}, A. and {Besenrieder}, J. and {Bhattacharyya}, W. and {Bigongiari}, C. and {Biland}, A. and {Blanch}, O. and {Bonnoli}, G. and {Bo{\v{s}}njak}, {\v{Z}}. and {Busetto}, G. and {Carosi}, R. and {Ceribella}, G. and {Cerruti}, M. and {Chai}, Y. and {Chilingarian}, A. and {Cikota}, S. and {Colak}, S.~M. and {Colin}, U. and {Colombo}, E. and {Contreras}, J.~L. and {Cortina}, J. and {Covino}, S. and {D'Amico}, G. and {D'Elia}, V. and {da Vela}, P. and {Dazzi}, F. and {de Angelis}, A. and {de Lotto}, B. and {Delfino}, M. and {Delgado}, J. and {Depaoli}, D. and {di Pierro}, F. and {di Venere}, L. and {Do Souto Espi{\~n}eira}, E. and {Dominis Prester}, D. and {Donini}, A. and {Dorner}, D. and {Doro}, M. and {Elsaesser}, D. and {Fallah Ramazani}, V. and {Fattorini}, A. and {Ferrara}, G. and {Foffano}, L. and {Fonseca}, M.~V. and {Font}, L. and {Fruck}, C. and {Fukami}, S. and {Garc{\'\i}a L{\'o}pez}, R.~J. and {Garczarczyk}, M. and {Gasparyan}, S. and {Gaug}, M. and {Giglietto}, N. and {Giordano}, F. and {Gliwny}, P. and {Godinovi{\'c}}, N. and {Green}, D. and {Hadasch}, D. and {Hahn}, A. and {Heckmann}, L. and {Herrera}, J. and {Hoang}, J. and {Hrupec}, D. and {H{\"u}tten}, M. and {Inada}, T. and {Inoue}, S. and {Ishio}, K. and {Iwamura}, Y. and {Jouvin}, L. and {Kajiwara}, Y. and {Karjalainen}, M. and {Kerszberg}, D. and {Kobayashi}, Y. and {Kubo}, H. and {Kushida}, J. and {Lamastra}, A. and {Lelas}, D. and {Leone}, F. and {Lindfors}, E. and {Lombardi}, S. and {Longo}, F. and {L{\'o}pez}, M. and {L{\'o}pez-Coto}, R. and {L{\'o}pez-Oramas}, A. and {Loporchio}, S. and {Machado de Oliveira Fraga}, B. and {Maggio}, C. and {Majumdar}, P. and {Makariev}, M. and {Mallamaci}, M. and {Maneva}, G. and {Manganaro}, M. and {Mannheim}, K. and {Maraschi}, L. and {Mariotti}, M. and {Mart{\'\i}nez}, M. and {Mazin}, D. and {Mender}, S. and {Mi{\'c}anovi{\'c}}, S. and {Miceli}, D. and {Miener}, T. and {Minev}, M. and {Miranda}, J.~M. and {Mirzoyan}, R. and {Molina}, E. and {Moralejo}, A. and {Morcuende}, D. and {Moreno}, V. and {Moretti}, E. and {Munar-Adrover}, P. and {Neustroev}, V. and {Nigro}, C. and {Nilsson}, K. and {Ninci}, D. and {Nishijima}, K. and {Noda}, K. and {Nozaki}, S. and {Ohtani}, Y. and {Oka}, T. and {Otero-Santos}, J. and {Palatiello}, M. and {Paneque}, D. and {Paoletti}, R. and {Paredes}, J.~M. and {Pavleti{\'c}}, L. and {Pe{\~n}il}, P. and {Perennes}, C. and {Persic}, M. and {Prada Moroni}, P.~G. and {Prandini}, E. and {Priyadarshi}, C. and {Puljak}, I. and {Rhode}, W. and {Rib{\'o}}, M. and {Rico}, J. and {Righi}, C. and {Rugliancich}, A. and {Saha}, L. and {Sahakyan}, N. and {Saito}, T. and {Sakurai}, S. and {Satalecka}, K. and {Schleicher}, B. and {Schmidt}, K. and {Schweizer}, T. and {Sitarek}, J. and {{\v{S}}nidari{\'c}}, I. and {Sobczynska}, D. and {Spolon}, A. and {Stamerra}, A. and {Strom}, D. and {Strzys}, M. and {Suda}, Y. and {Suri{\'c}}, T. and {Takahashi}, M. and {Tavecchio}, F. and {Temnikov}, P. and {Terzi{\'c}}, T. and {Teshima}, M. and {Torres-Alb{\`a}}, N. and {Tosti}, L. and {Truzzi}, S. and {van Scherpenberg}, J. and {Vanzo}, G. and {Vazquez Acosta}, M. and {Ventura}, S. and {Verguilov}, V. and {Vigorito}, C.~F. and {Vitale}, V. and {Vovk}, I. and {Will}, M. and {Zari{\'c}}, D. and {Jorstad}, S.~G. and {Marscher}, A.~P. and {Boccardi}, B. and {Casadio}, C. and {Hodgson}, J. and {Kim}, J. -Y. and {Krichbaum}, T.~P. and {L{\"a}hteenm{\"a}ki}, A. and {Tornikoski}, M. and {Traianou}, E. and {Weaver}, Z.~R.},
        title = "{H.E.S.S. and MAGIC observations of a sudden cessation of a very-high-energy {\ensuremath{\gamma}}-ray flare in PKS 1510-089 in May 2016}",
      journal = {\aap},
         year = 2021,
        month = apr,
       volume = {648},
          eid = {A23},
        pages = {A23},
          doi = {10.1051/0004-6361/202038949},
archivePrefix = {arXiv},
       eprint = {2012.10254},
 primaryClass = {astro-ph.HE},
       adsurl = {https://ui.adsabs.harvard.edu/abs/2021A&A...648A..23H}
}

@ARTICLE{cerruti20,
       author = {{Cerruti}, Matteo},
        title = "{Leptonic and Hadronic Radiative Processes in Supermassive-Black-Hole Jets}",
      journal = {Galaxies},
         year = 2020,
        month = oct,
       volume = {8},
       number = {4},
        pages = {72},
          doi = {10.3390/galaxies8040072},
archivePrefix = {arXiv},
       eprint = {2012.13302},
 primaryClass = {astro-ph.HE},
       adsurl = {https://ui.adsabs.harvard.edu/abs/2020Galax...8...72C}
}

@ARTICLE{zhang+21,
       author = {{Zhang}, Hao and {Sironi}, Lorenzo and {Giannios}, Dimitrios},
        title = "{Fast Particle Acceleration in Three-dimensional Relativistic Reconnection}",
      journal = {\apj},
         year = 2021,
        month = dec,
       volume = {922},
       number = {2},
          eid = {261},
        pages = {261},
          doi = {10.3847/1538-4357/ac2e08},
archivePrefix = {arXiv},
       eprint = {2105.00009},
 primaryClass = {astro-ph.HE},
       adsurl = {https://ui.adsabs.harvard.edu/abs/2021ApJ...922..261Z}
}

@ARTICLE{zechlemoine21,
       author = {{Zech}, Andreas and {Lemoine}, Martin},
        title = "{Electron-proton co-acceleration on relativistic shocks in extreme-TeV blazars}",
      journal = {\aap},
         year = 2021,
        month = oct,
       volume = {654},
          eid = {A96},
        pages = {A96},
          doi = {10.1051/0004-6361/202141062},
archivePrefix = {arXiv},
       eprint = {2108.12271},
 primaryClass = {astro-ph.HE},
       adsurl = {https://ui.adsabs.harvard.edu/abs/2021A&A...654A..96Z}
}

@ARTICLE{park+21,
       author = {{Park}, Jongho and {Hada}, Kazuhiro and {Nakamura}, Masanori and {Asada}, Keiichi and {Zhao}, Guangyao and {Kino}, Motoki},
        title = "{Jet Collimation and Acceleration in the Giant Radio Galaxy NGC 315}",
      journal = {\apj},
         year = 2021,
        month = mar,
       volume = {909},
       number = {1},
          eid = {76},
        pages = {76},
          doi = {10.3847/1538-4357/abd6ee},
archivePrefix = {arXiv},
       eprint = {2012.14154},
 primaryClass = {astro-ph.HE},
       adsurl = {https://ui.adsabs.harvard.edu/abs/2021ApJ...909...76P}
}

@ARTICLE{boccardi+21,
       author = {{Boccardi}, B. and {Perucho}, M. and {Casadio}, C. and {Grandi}, P. and {Macconi}, D. and {Torresi}, E. and {Pellegrini}, S. and {Krichbaum}, T.~P. and {Kadler}, M. and {Giovannini}, G. and {Karamanavis}, V. and {Ricci}, L. and {Madika}, E. and {Bach}, U. and {Ros}, E. and {Giroletti}, M. and {Zensus}, J.~A.},
        title = "{Jet collimation in NGC 315 and other nearby AGN}",
      journal = {\aap},
         year = 2021,
        month = mar,
       volume = {647},
          eid = {A67},
        pages = {A67},
          doi = {10.1051/0004-6361/202039612},
archivePrefix = {arXiv},
       eprint = {2012.14831},
 primaryClass = {astro-ph.HE},
       adsurl = {https://ui.adsabs.harvard.edu/abs/2021A&A...647A..67B}
}

%
%
%

\end{document}